\documentclass[aps,superscriptaddress,preprint,nofootinbib]{revtex4}

\usepackage{amsmath}
\usepackage{graphicx}
\usepackage{amssymb}
\usepackage{dcolumn}
\usepackage{bm}
\usepackage{color}

\makeatletter

\begin{document}

\title{Near and far-field hydrodynamic interaction of two chiral squirmers}

\author{Ruma Maity}
\affiliation{Department of Physics, Indian Institute of Technology Kharagpur, Kharagpur 721302, India}

\author{P. S. Burada} \thanks{\tt Corresponding author:psburada@phy.iitkgp.ac.in}
\affiliation{Department of Physics, Indian Institute of Technology Kharagpur, Kharagpur 721302, India}

\date{\today}

\begin{abstract}

Hydrodynamic interaction strongly influences the collective behavior of the microswimmers. With this work, we study the behavior of two hydrodynamically interacting self-propelled chiral swimmers in the low Reynolds number regime, considering both the near and far-field interactions. We use the chiral squirmer model, a spherically shaped body with non-axisymmetric surface slip velocity, which generalizes the well-known squirmer model. We calculate the lubrication force between the swimmers when they are very close to each other. By varying the slip coefficients and the initial configuration of the swimmers, we investigate their hydrodynamic behavior. In the presence of lubrication force, the swimmers either repel each other or exhibit bounded motion where the distance between the swimmers alters periodically. 
The lubrication force favors the bounded motion in some parameter regime. 
This study is helpful to understand the collective behavior of dense suspension of ciliated microorganisms and artificial swimmers.

\end{abstract}

\maketitle

\section{Introduction}

The swimming behavior of microorganisms is different from that of the macroworld \cite{purcell}. In the former case, viscous forces dominate over the inertia of the body. This belongs to the low Reynolds number swimming \cite{purcell,happel}. Different microorganisms employ various propulsion mechanisms, e.g., 
{\it Escherichia coli} use run and tumble mechanism to propel in a fluid \cite{Larsen}, ciliated microorganisms swim with the help of the metachronal waves generated by the synchronous beating of cilia \cite{lighthill, blake} and sperm cells move with the flagella attached to its body \cite{friedrich}. 
To understand the propulsion mechanism of microswimmers, 
various models are available in the literature 
\cite{lighthill, blake, friedrich, purcell2, jiang, Eric_book}. 
Though the microorganisms are smaller in size, collectively they can influence the climate and human life in various ways. 
For example, massive plankton blooms in the ocean, harmful red tides along the coastline, bioconvection \cite{pallat}, nutrient uptake \cite{kirchman}, 
active turbulence \cite{dunkel}, and they may even influence the viscosity of the surrounding medium in which they swim \cite{sokolov, haines}. 
In the past, the suspension of microswimmers was studied using a continuum model \cite{fisham, pedley, metcalfe, saintillan} which works well for dilute suspensions only and generally not applicable for larger cell concentrations. 
For a denser system, near field interactions are vital. 
The hydrodynamic interaction among the miroswimmers has been extensively studied both experimentally \cite{ishikawa, drescher, Goldstein_JFM1, Goldstein_JFM2, darnton} and theoretically \cite{ishikawa, simmonds, ishikawa2, gotze, pooley, molina, burada}. 
Some of the former studies are devoted to the two swimmers system \cite{ishikawa, drescher, Goldstein_JFM1, Goldstein_JFM2, simmonds, pooley, burada}. 
Indeed, all these studies take into account pure hydrodynamic interaction among the microswimmers. When the swimmers are far away from each other, the interaction among them can be expressed in terms of a multipole expansion. Conversely, while the swimmers are very close to each other, one needs to use the lubrication theory to calculate their near-field interaction. 

Several studies on the near and far field hydrodynamic interaction between two or more axisymmetric swimmers \cite{ishikawa,simmonds} are available where the swimmers change their direction of movement exhibiting attractive or repulsive behavior depending upon their respective velocity field strengths. 
A popular squirmer model \cite{lighthill, blake} 
has been used to understand the hydrodynamic interactions among the swimmers.
However, the squirmer model has its own limitation as it can be associated only with the translational motion of the body. 
Consequently, the direction of motion of the body changes either due to the rotational diffusion or the hydrodynamic interaction with another squirmer. 
In general, many microorganisms are able to change their direction of movement by rotation of the orientational vector giving rise to helical motion \cite{crenshaw}. Henceforth, the chiral squirmer model which is a generalization of the 
squirmer model \cite{burada, ruma} is more applicable to study 
the collective behaviour of the swimmers. 
In the latter model, the tangential slip velocity on the surface of a non-deformable spherical body is non-axisymmetric, and as a result the chiral flows and helical paths can be generated by the swimmer. 

Similar to simple squirmers, it has been reported that a pair of chiral squirmers also portray various behaviours, e.g., monotonic divergence, divergence, monotonic convergence, convergence and even a bounded state \cite{burada, mirzakhanloo, theers} as a result of their mutual hydrodynamic interaction. 
The helical propulsion of the swimmers leads to this peculiar bounded state, where the swimmers periodically come closer to and go distant apart from each other periodically, this was reported in our earlier work \cite{burada}.
In the latter study, only the far-field hydrodynamic interaction was considered for simplicity and ignored the lubrication force which arises when the swimmers are very close to each other \cite{burada}.

In this article, we study the combined behavior of two chiral swimmers 
considering both the near and far-field hydrodynamic interactions. 
We compute the lubrication force between two swimmers when they approach very close to each other. Further, we investigate the complete hydrodynamic behavior of two swimmers. 
The paper is organized as follows. 
The general chiral squirmer model is briefly discussed in section~\ref{sec:model}. 
The lubrication force between two swimmers is calculated in section~\ref{sec:lub}.
The hydrodynamic interaction, both in the near and far fields, between two swimmers is discussed in section~\ref{sec:States}. 
Influence of initial conditions of swimmers on their hydrodymic behavior is explained in 
section~\ref{sec:ini_pos}.
The main conclusions are provided in section~\ref{sec:conclusions}.

\section{The chiral squirmer model}
\label{sec:model}

\begin{figure}[t]
\centering
\includegraphics[scale=0.75]{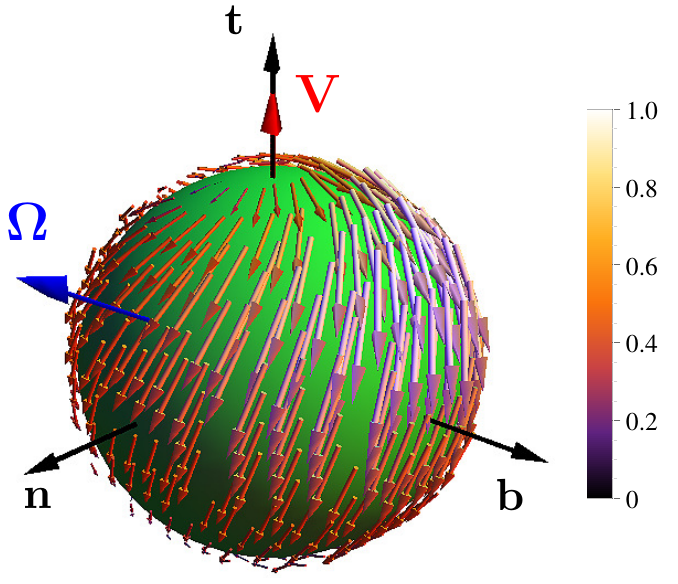}
\caption{Example of surface slip velocity patterns of a chiral squirmer in the body-fixed reference frame {\bf n}, {\bf b}, and ${\bf t}$.
Here, we set the velocity and rotation rate of the swimmer as 
${\bf V} = v(0,0,1)$ and $ {\bf \Omega} = v(1/\sqrt{2}, 0, 1/\sqrt{2})/a$, respectively. 
The slip coefficients of the second mode are chosen as 
$\beta_{2 0}^r = v/3$, $\gamma_{2 0}^r = v/3$, and the others are zero.}
\label{fig:slip}
\end{figure}

The low Reynolds number swimmers obey the Stokes equation \cite{happel},
\begin{equation}
 \eta\nabla^2 \mathbf{u} = \nabla p\,,
 \label{eq:stokes}
\end{equation}
where $\eta$ is the viscosity, ${\bf u}$ is the velocity field, 
and $p$ is the pressure field which plays the role of a Lagrange multiplier to impose the incompressibility constraint $ {\bf \nabla} \cdot {\bf u} = 0$.
A chiral squirmer is a rigid spherical body of radius $a$. 
On its surface, we prescribe a surface slip velocity ${\bf S} (\theta,\phi)$ which is tangential to the surface and parameterized by the polar and azimuthal angles $\theta$ and $\phi$, respectively, in a body-fixed frame defined by three orthogonal unit vectors attached to the sphere center ${\bf n}$, ${\bf b}$, and ${\bf t}$ (see Fig.~\ref{fig:slip}).  
It is convenient to express this surface slip pattern using gradients of spherical harmonics that form a basis for tangential vectors on the surface \cite{happel}. 
The slip velocity can then be expressed in the form \cite{ruma,burada}
\begin{align}
\label{eq:slip}
{\bf S}(\theta, \phi) & = \sum_{l=1}^{\infty} \sum_{m= -l}^l
\Big[-\beta_{lm}\, {\boldsymbol \nabla}_s 
\left( P_l^m(\cos\theta) \,e^{i m \phi} \right) \nonumber \\  
& + \gamma_{lm}\, {\bf \hat{r}} \times {\boldsymbol \nabla}_s \left( P_l^m(\cos\theta) \,e^{i m \phi} \right) \Big]\,, 
\end{align}
where ${\boldsymbol \nabla}_s$ is the gradient operator on the surface of the sphere defined as 
${\boldsymbol \nabla}_s = {\bf e}_\theta \, {\partial}/{\partial}\theta +(1/\sin\theta)\, {\bf e}_\phi {\partial}/{\partial}\phi$, 
${\bf \hat{r}}$ is the unit vector in radial direction, 
$P_l^m(\cos\theta) \,e^{i m \phi}$
are non-normalized spherical harmonics, where $P_l^m(\cos\theta)$ denotes 
Legendre polynomials.
The complex coefficients $\beta_{l m}$ and $\gamma_{l m}$ are the mode amplitudes of the prescribed surface slip velocity. 
We introduce the real and imaginary parts of these amplitudes 
as $\beta_{l m} = \beta_{l m}^r + i\,m\, \beta_{l m}^i$ and $\gamma_{l m} = \gamma_{l m}^r + i\,m\, \gamma_{l m}^i$ 
with complex conjugates $\beta_{l m}^\ast = (-1)^m \beta_{l, -m}$ and $\gamma_{l m}^\ast = (-1)^m \gamma_{l, -m}$, respectively.

\begin{figure}
\includegraphics[scale=0.75]{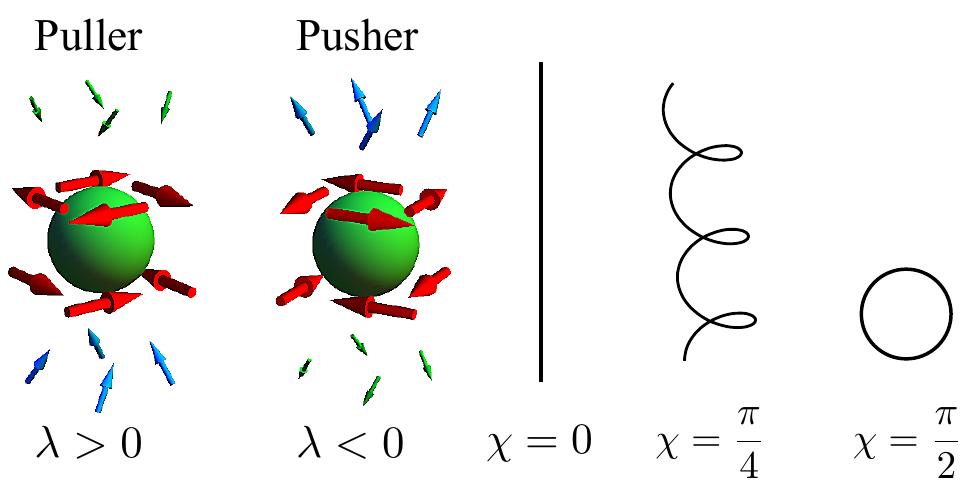}
\caption{
Chiral flow pattern exhibited by a pusher and a puller type chiral squirmer in three dimensions, for $\lambda = 3 \beta^A_{20} v = 3 \gamma^A_{20} v$ in Eq. ~\ref{eq:VelocityField}, 
and the chiral squirmer's path for different angles $\chi$ between the velocity ${\mathbf{V}}$ and the rotation rate ${\boldsymbol \Omega}$. 
The initial velocities and rotation rates of both the swimmer are set 
to $v(0,0,1)$ and $v(\sin \chi, 0, \cos \chi)/a$, respectively. 
For the flow patterns we set $\chi = \pi/4$ and $\lambda = v$ for puller, and 
$\chi = \pi/4$ and $\lambda = -v$ for pusher.}
\label{fig:Sketch2}
\end{figure}

The velocity $\bm{V}$ and the rotation rate $\bm{\Omega}$ of the swimmer can be determined directly using the surface slip velocity Eq.~\ref{eq:slip} \cite{stone}.
They can be expressed in the body fixed reference frame as 
${\bf V} = 2(\beta_{11}^r , \, \beta_{11}^i , \, \beta_{10}^r )/3$ and
${\boldsymbol \Omega} = (\gamma_{11}^r, \, \gamma_{11}^i, \, \gamma_{10}^r )/a$,
respectively.
Without loss of generality, the body-fixed reference frame $({\bf n}, {\bf b}, {\bf t})$ can be chosen such that ${\bf t}$ points in the direction of motion.
Accordingly, we have $\beta_{11}^r = \beta_{11}^i = 0$ and we write $\beta_{1 0}^r = 3 v/2$ such that $v=|{\bf V}|$ is the speed of the swimmer.
Thus, the velocity and rotation rate of the chiral swimmer read \cite{burada},
\begin{align}
{\bf V} & = v \,  {\bf t} \ , \\ 
\label{eq:vel_rot}
{\bf \Omega} & =  \frac{\gamma_{11}^r}{a} \,{\bf n} +  \frac{\gamma_{11}^i}{a} \,{\bf b} +  \frac{\gamma_{10}^r}{a} \, {\bf t} \,.
\end{align} 
Also, for simplicity, we choose that the swimmer has rotation rate in the ${\bf n-t}$ plane only. With this choice, we have $\gamma_{11}^i = 0$.
In addition, we choose the magnitude of the rotation as 
$|{\boldsymbol \Omega}| = v/a$ such that the components of the rotation rate 
are expressed as 
$\gamma_{11}^r/a = (v/a)\sin \chi \,\, , \gamma_{11}^i/a = 0$ and $\gamma_{10}^r/a = (v/a)\cos \chi$, where $\chi$ is the angle between 
${\mathbf{V}}$ and ${\boldsymbol \Omega}$. 
The corresponding flow field and the pressure field of the swimmer can be obtained by solving Eq.~\ref{eq:stokes} with the surface slip (Eq.~\ref{eq:slip}) 
in the lab frame of reference. They read \cite{burada},
\begin{align}
\label{eq:VelocityField}
{\bf u}_\mathrm{lf}({\bf r})  
   & = \frac{3 v}{2} \frac{a^3}{r^3} 
   \left[ P_1({\bf t} \cdot {\bf \hat{r}})\,{\bf \hat{r}} - \frac{\bf t}{3} 
   \right]   
+ 3\,\beta_{2 0}^r \left( \frac{a^4}{r^4} - \frac{a^2}{r^2} \right)\,
P_2({\bf t} \cdot {\bf \hat{r}})\, {\bf \hat{r}}  \nonumber \\
& + \beta_{2 0}^r \frac{a^4}{r^4} 
P_2^{\prime}\left( {\bf t} \cdot {\bf \hat{r}} \right) [ ({\bf t} \cdot {\bf \hat{r}}) {\bf \hat{r}} - {\bf t} ] 
- \gamma_{2 0}^r\,\frac{a^3}{r^3} \, 
P_2^{\prime}\left( {\bf t} \cdot {\bf \hat{r}} \right)
{\bf t} \times {\bf \hat{r}} \,,\\ 
\label{eq:pressure_LB}
  p_\mathrm{lf}({\bf r})
  & =  - 2 \eta\,\beta_{2 0}^r \,\frac{a^2}{r^3} \,P_2\left({\bf t} \cdot {\bf \hat{r}}\right) \,,
\end{align}
where 
${\bf t}$ is the swimming direction, 
$r$ is the distance from the center of the swimmer where the flow field is determined, 
${\bf \hat{r}} = {\bf r}/r$ is the radial vector,
$P_2(x)$ denotes a second-order Legendre polynomial, and 
$P_2^{\prime} = dP_2/dx$ with $x = \mathbf{t}\cdot \mathbf{\hat{r}} = \cos \theta$.
Note that in Eq.~(\ref{eq:VelocityField}) the higher order terms $l > 2$ are being ignored as their contribution is negligible in the current study.
To have a minimal model, in Eq.~(\ref{eq:VelocityField}) we have ignored $l = 2$ modes with $m\neq 0$. 
However, it is straightforward to include the additional terms in the analysis.
Depending on the sign of the ratio $\beta = \beta_{2 0}^r/\beta_{1 0}^r$, the swimmer can be classified as a puller (for $\beta > 0$) or pusher (for $\beta < 0$) type 
(see Fig.~\ref{fig:Sketch2}). 
While pullers have an extensile force dipole, resulting, e.g., from the front part of the body, pushers have a contractile force dipole arising, e.g., from the rear part of the body \cite{Eric_book}, see Fig.~\ref{fig:Sketch2}. 
Note that the flow field in the body frame (bf) can be obtained from that in the lab frame (lf) as 
${\bf u}_\mathrm{bf}({\bf r}) = {\bf u}_\mathrm{lf}({\bf r}) - {\bf V} - {\boldsymbol \Omega} \times {\bf r}$.

The equations of motion of the swimmer can be obtained 
using the force and torque balance conditions \cite{kim}. 
They read,
\begin{align}
\dot{\textbf{q}} = \textbf{V}, \,\,\,\, 
\dot{\textbf{n}} = {\boldsymbol \Omega} \times \textbf{n},\,\,\,\,
\dot{\textbf{b}} = {\boldsymbol \Omega} \times \textbf{b},\,\,\,\,
\dot{\textbf{t}} = {\boldsymbol \Omega} \times \textbf{t}\,,
\label{eqn:single}
\end{align}
where $\textbf{q}$ is the position of the swimmer in the lab frame of reference 
and dot represents the derivative with respect to time. 
For ${\mathbf{V}} \parallel {\boldsymbol \Omega}$, we get $\chi = 0$, and 
the resulting swimming path of the swimmer is a straight line. 
In this case, the swimmer rotates around the axis of motion. 
For $\chi = \pi/2$, the swimmer moves in a circular path in a plane. 
For other values of $\chi$, the path is a helix 
(see Fig.~\ref{fig:Sketch2}) \cite{burada}. 
Note that Eq.~\ref{eqn:single} determine the motion of a single isolated squirmer, whereas for a pair of squirmers we need to take into account the hydrodynamic interaction between them which we study in the following.

\section{Lubrication force between two chiral swimmers}
\label{sec:lub}

\begin{figure}
\includegraphics[scale=0.5]{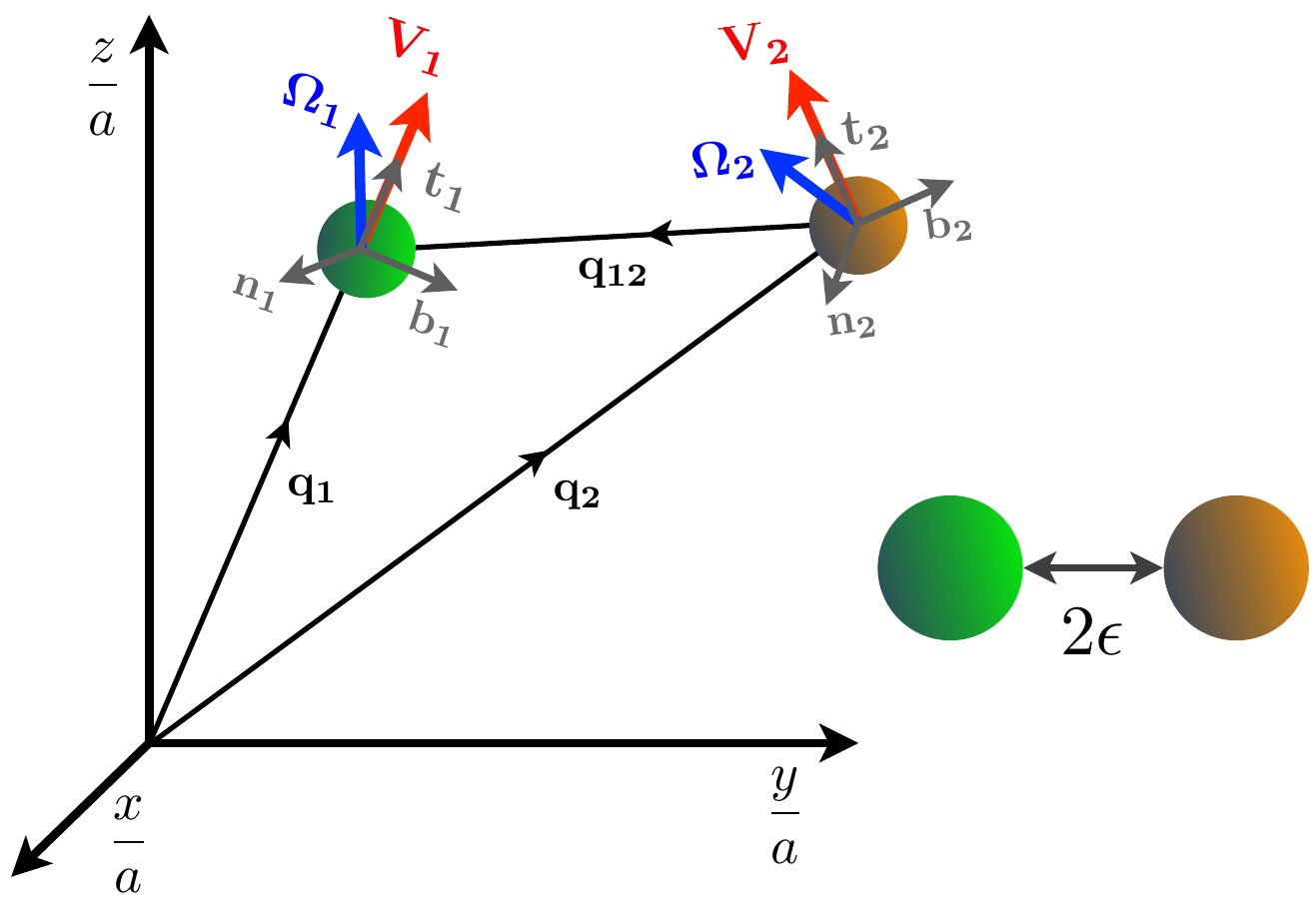}
\caption{
Schematic representation of two chiral squirmers in the laboratory frame of reference. 
When swimmers are close to each other, i.e., the distance between them 
$R = |\textbf{q}_{ij}| = r \le 2(a+ \epsilon)$, where $a$ is the radius of the swimmer, 
then the lubrication forces control the hydrodynamic behavior of the swimmers. 
Otherwise, the far field hydrodynamic interactions dominate.}
\label{fig:sketch1}
\end{figure}

A substantial work has been done in low Reynolds number swimming near an air-liquid interface \cite{leonardo,trouilloud,wang,di}. 
To find the force on the body near the air liquid interface considering perfect slip, mirror image technique has been used \cite{wang,di}. 
To calculate the lubrication force between two swimmers a similar approach can be 
adapted. 
Here, in place of image, both the swimmers are real and their dynamics is controlled by the Stokes equation. 
The lubrication force acting on a swimmer can be calculated by knowing the velocity field 
of the nearby swimmer, see for the details provided in appendix \ref{lub_cal}.
The component of the lubrication force acting on a swimmer along its swimming direction 
reads,
\begin{align}
\label{eq:lub_force}
F_{_Z} & \approx \frac{3 \pi B a^2}{2} \ln \epsilon \,
\end{align}
where $B = \beta_{10}^1 t_{13} - \beta_{10}^2 t_{23}$, 
$\epsilon$ is half the distance between the swimmers, 
$t_{13}= \mathbf{t}_1 \cdot \mathbf{e}_{_Z}$, 
$t_{23} = \mathbf{t}_2 \cdot \mathbf{e}_{_Z}$,
$\mathbf{e}_{_Z}$ is the unit vector along the $Z$ direction,
$\mathbf{t}_1$ and  
$\mathbf{t}_2$ are the orientation vectors of the swimmer one and two, respectively 
(for details see the appendix \ref{lub_cal}).
Taking into account the solution for squeezing motion of two rigid spheres, 
we can find the velocity of axisymmetric squirmer in the lubrication region 
as $U_{_Z} \sim \epsilon \log\epsilon$ \cite{wang, yoshinaga}. 
Similarly, the velocity of the chiral squirmer can be obtained as 
\begin{align}
\label{eq:lub_vel}
U_{_Z} = -2 a^2 B \epsilon \ln \epsilon \,.
\end{align}
Also note that the lubrication torques are of the order $O(\epsilon^{1/2})$, 
and which can be neglected in the limit $\epsilon \to 0$. 

Notably, the results obtained here agree with the ones by Wang \textit{et al.}\cite{wang}. However, the later is the case of axisymmetric squirmers, whereas the present study deals with chiral squirmers. The flow field in the narrow gap between the axisymmetric squirmers contains only radial and polar components. However, for the chiral squirmers,  the flow field in the lubrication region contains an azimuthal component in addition to the radial and polar components. 
Note that the lubrication force acting on the case of axisymmetric squirmers contains the polar slip coefficients only. On the other hand, the lubrication torque contains the azimuthal slip coefficients for a chiral squirmer. 
However, contribution from the lubrication torque is insignificant in the hydrodymic interaction of chiral squirmers. 
Consequently, the calculated lubrication forces are the same for both axisymmetric and chiral squirmers despite having different flow fields.

In the presence of the lubrication force, 
the corresponding equations of motion of the swimmers are given by, 
\begin{align}
\label{eq:gg}
{\bf \dot{q}}_i
& = {\bf U}_i + \mathbf{U}^{\mathrm{lub}}_{i} +\sum^2\limits_{\substack{j=1 \,;\, i\neq j}} \mathbf{u}_j(\mathbf{q}_{ij}, {\bf n}_2, {\bf b}_2, {\bf t}_2) \, \nonumber \\
\left[\begin{array}{c}  {\bf \dot{n}}_i  \\ {\bf \dot{b}}_i \\{\bf \dot{t}}_i \end{array}\right]  
& = \left[{\boldsymbol \Omega}_i+ \sum^2\limits_{\substack{ i\neq j \\ j = 1}}\bm{\omega}_j(\mathbf{q}_{ij}, {\bf n}_2, {\bf b}_2, {\bf t}_2) \right] 
\times   
\left[\begin{array}{c} {\bf n}_i  \\ {\bf b}_i \\ {\bf t}_i \end{array}\right],
\end{align}
where $\mathbf{U}^\mathrm{lub}_i = U_{_Z}(\cos \theta^\prime \mathbf{t}_i - \sin \theta^\prime \mathbf{n}_i)$ is the additional velocity contribution arising due to the  other swimmer in the lubrication region, 
$\theta^\prime = \cos^{-1}(\mathbf{t}_i \cdot \mathbf{e}_{_Z})$, and the 
vorticity field $\boldsymbol{\omega} = ({\boldsymbol \nabla} \times {\bf u}) /2$. 
Note that $\mathbf{U}^{\mathrm{lub}}_i = 0$ for $R > 2(a + \epsilon)$, and 
$\mathbf{U}_i + \sum^2\limits_{\substack{j=1 \,;\, i\neq j}} \mathbf{u}_j = \sum^2\limits_{\substack{j=1 \,;\, i\neq j}}\bm{\omega}_j= 0$ for 
$R \le 2(a+ \epsilon)$, where 
$\epsilon \ll a$ and $R = |\textbf{q}_{ij}|$ is the radial distance between the squirmers.

\section{Hydrodynamic behavior of two chiral swimmers}
\label{sec:States} 

To study the hydrodymic interaction between two swimmers, 
we numerically solve the Eq.~(\ref{eq:gg}), to calculate the trajectories of 
two chiral swimmers and investigate their combined behavior.
For simplicity, we consider chiral swimmers having translational velocities of equal magnitudes, i.e., $|\mathbf{V}_1| = |\mathbf{V}_2| = v$. The rotation rates of the swimmers are in general different and read, 
${\bf \Omega_1} = v(\sin\chi_1, 0,\cos\chi_1)/a$ for swimmer one and 
${\bf \Omega_2} = v(\sin\chi_2, 0,\cos\chi_2)/a$ for swimmer two. 
Note that changes in $\chi_1$ and $\chi_2$ modify the corresponding torsion and curvature of the swimmers' helical trajectories. 
Also, $l > 1$ modes in the velocity field Eq.~(\ref{eq:VelocityField}) 
play a crucial role in the hydrodynamic interaction between the swimmers. 
As mentioned earlier, we consider up to $l = 2$ modes in the flow field. 
We choose $l = 2$ modes corresponding to swimmer one as 
$3\beta_{20}^{r} = 3\gamma_{20}^{r} = \lambda_1$ and similarly for swimmer two as 
$3\beta_{20}^{r} = 3\gamma_{20}^{r} = \lambda_2$.
Note that for $\lambda_1 \neq \lambda_2$, the swimmers differ in their chiral flows that they generate.
Thus, variation in $\chi_i$ and $\pm \lambda_i \,(i = 1,2)$ determine the nature of the interaction between the swimmers and gives rise to several interesting swimming characteristics.
As mentioned earlier, the sign of $\lambda_i$ 
(see Eq.~\ref{eq:VelocityField}) decides the nature of the swimmer, i.e. pusher or puller type. Accordingly, we consider the sub cases, i.e., 
pusher - pusher: $(-\lambda_1, -\lambda_2)$, 
puller - puller: $ (\lambda_1, \lambda_2)$, 
pusher - puller: $(-\lambda_1, \lambda_2)$, and
puller - pusher: $ (\lambda_1, -\lambda_2)$. 
We have considered various possible initial configurations for the swimmers. 
Out of all, in this paper, we present only the planar configuration, 
where both the swimmers start initially on the $xy$-plane, 
by a distance $R_0$, moving in the positive $z$-direction.  
In the planar configuration, swimmers get enough time to interact with each other, whereas it may not be the case in other configurations.
This particular choice of the configuration recovers the known behaviors exhibited by two simple squirmers (without chirality) 
and some additional exciting behaviors discussed below. 

\begin{figure}[t!]
\includegraphics[scale=0.75]{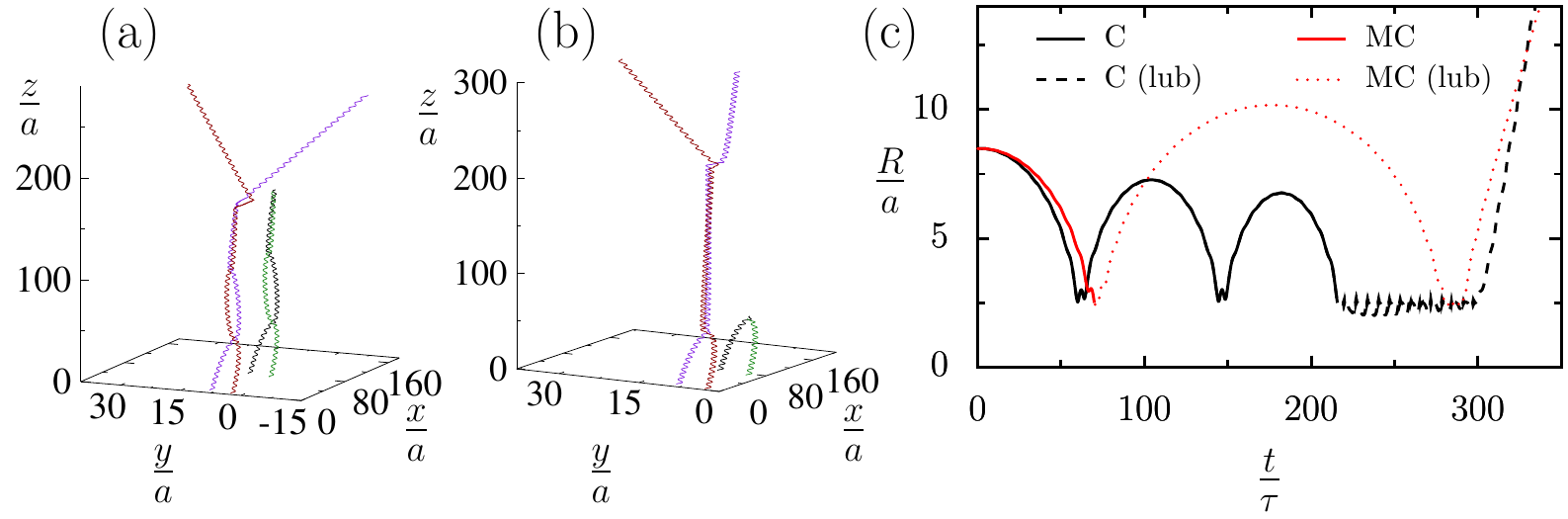}
\caption{
Panel $(a)$ shows trajectories of two converging (C) swimmers in the absence of 
near-field interaction (green and black helices shifted by 70 units along $x$-direction) and in presence of it (Purple and red). Similarly, panel $(b)$ shows trajectories of two monotonically converging (MC) swimmers in the absence and in the presence of near field interactions. Panel $(c)$ shows
the corresponding distance $R$ between two hydrodynamically interacting swimmers as a function of time both in absence and in presence of the lubrication force. Here, lengths are scaled by radius of the swimmer $a$ and 
time scaled by $\tau = v/a$.
Note that lubrication forces convert the monotonic convergence (MC) and convergence states  (C) into divergence state (D). 
Here, for the state C, we choose 
$\chi_1 = \chi_2 = \pi/6$, $\lambda_1 = -2.5$ and $\lambda_2 = 2.5$.
For MC, we choose $\chi =  \pi/6$, $\lambda_1 = -2$ and $\lambda_2 = 2$.}
\label{fig:lub2}
\end{figure}

Note that the hydrodynamic interaction between two chiral swimmers in the far-field limit has been explored in the previous work \cite{burada}. Five different swimming states were observed- $(i)$ bounded (B), in which the swimmers oscillate around an average trajectory, $(ii)$ monotonic divergence (MD), in which the swimmers drift away from each other from the beginning, $(iii)$ divergence (D), in which initially the swimmers are attracted to each other but in the long time limit they move away from each other due to the growing repulsion between them, $(iv)$ monotonic convergence (MC), in which the swimmers due to strong attraction monotonically approach each other at a distance where near-field interaction is crucial than the far-field interaction, and $(v)$ convergence (C), in which the swimmers initially oscillate about an average trajectory and then converge towards each other due to the hydrodynamic attraction between them. 
However, if the near-field interactions are dominant as in the case of a dense suspension, then the fate 
of the last two states, say, MC and C, were unknown. 
In this work, as mentioned earlier, we consider both the near and far field interactions to get the complete hydrodymic behavior of two chiral swimmers. 
Fig.~\ref{fig:lub2} depicts the behavior of two chiral swimmers which exhibit attractive behavior in the absence and in the presence of the lubrication forces. 
Due to the lubrication force, the C and MC states are converted into the D state, see Fig.~\ref{fig:lub2}. 
Note that the lubrication force is repulsive in nature. As a result, as the swimmers approach each other, i.e. as $R \le 2(a + \epsilon)$ (see Eq.~\ref{eq:gg}), they start to repel each other and diverge. 

\begin{figure*}[htb!]
\includegraphics[scale=0.675]{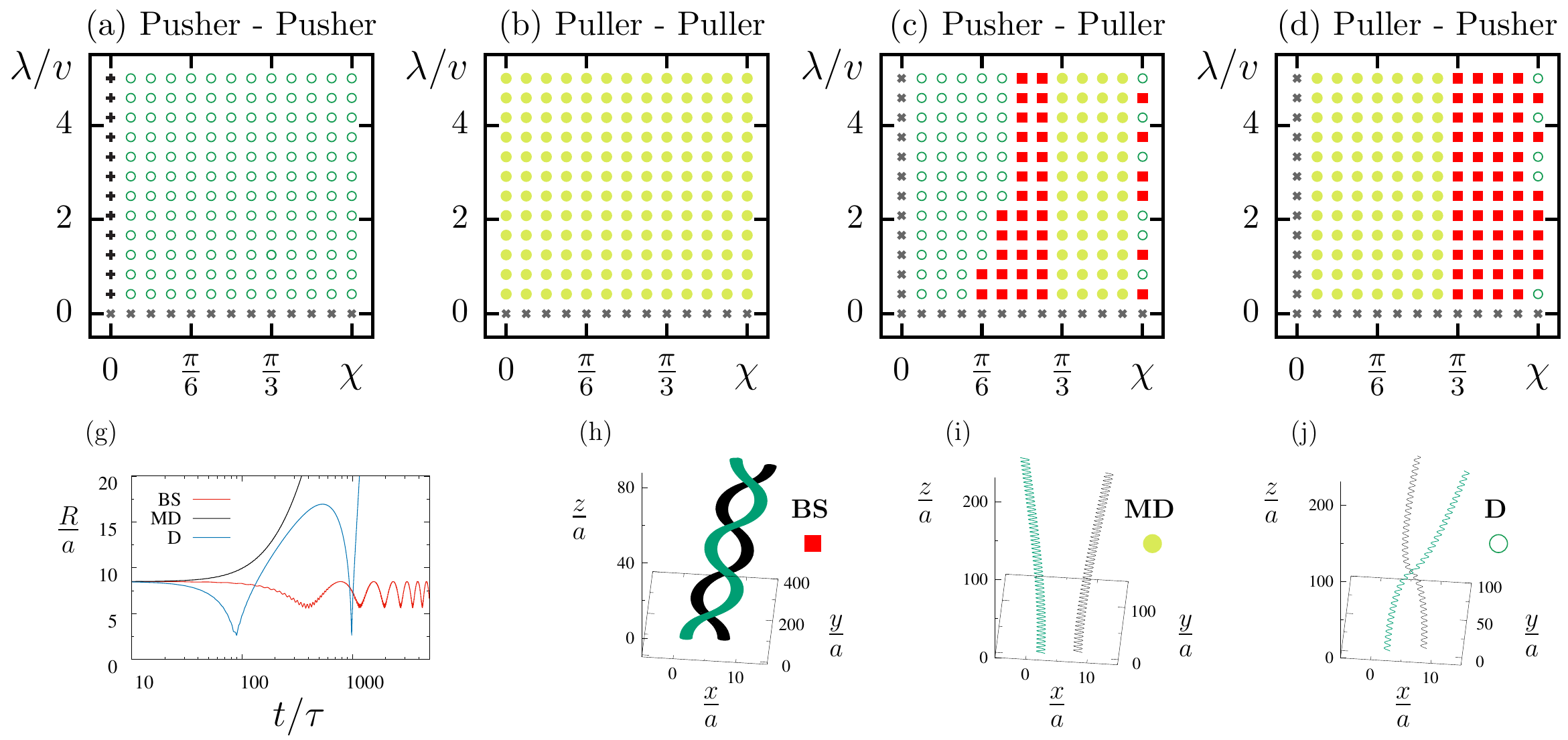}
\caption{
(a)-(d) Swimming states of two hydrodynamically interacting chiral swimmers of various combinations, i.e., pusher or puller. 
Hollow circles - divergence (D), solid circles- monotonic divergence (MD), squares- bounded (B), cross- parallel swimming and plus- forbidden states. 
(e) The corresponding distance $R$ between the swimmers 
is plotted as a function of time $t$. 
(f)- (h) Swimming trajectories corresponding to different states, 
for the values $\chi = \pi/3$ and $(\lambda_1, \lambda_2) = v(1, -1)$ for BS, 
$\chi = \pi/3$ and $(\lambda_1, \lambda_2) = v(-1, 1)$ for MD, 
and $\chi = 5 \pi/12$ and $(\lambda_1, \lambda_2) = v(-1, -1)$ for D.
Here, lengths are scaled by radius of the swimmer $a$, 
time scaled by $\tau = v/a$, and velocity with $v$.
The initial position of swimmer one is $(9,9,0)a$ and for the swimmer two is $(3,3,0)a$. 
The initial velocities and rotation rates of both the swimmers are set 
to $v(0,0,1)$ and $(v/a)(\sin \chi, 0, \cos \chi)$, respectively.}
\label{fig:lam-chi}
\end{figure*}

Fig.~\ref{fig:lam-chi} shows the hydrodymic behavior of two identical chiral swimmers. 
Here, we set $\chi_1 = \chi_2 = \chi$ and $|\lambda_1| = |\lambda_2| = \lambda$, i.e., the relative orientations of the swimmers with respect to their motion and 
the strength of the hydrodynamic flow fields of both the swimmers are identical. 
As mentioned earlier, due to the lubrication force which is repulsive in nature, 
MC and C states do not survive, leaving mainly D, MD, and B states in the state diagrams. 
Note that for the pusher--pusher combination, for $\chi = 0$, we obtain forbidden states (black plus). Two pushers swimming in parallel lines attract each other and may converge to a locked state, considered as a numerical artifact. However, for $\chi \neq 0$, the pushers move in a helical path, and the locked state does not appear. 
Note that we can extend this work for non-identical swimmers, i.e., the parameters $\lambda_1,\lambda_2,\chi_1$, and $\chi_2$ can be varied to study the hydrodymic behavior of two chiral swimmers. See fig.~\ref{fig:states} in the appendix~\ref{state} for more details.

\begin{figure}[htb!]
\centering
\includegraphics[scale=0.75]{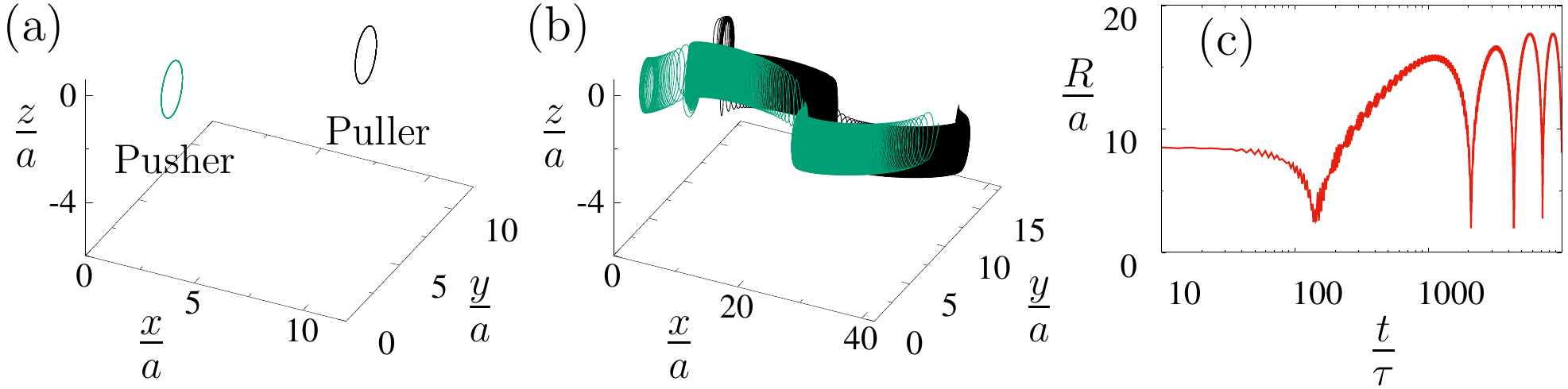}
\caption{
$(a)$ Circular motion of two swimmers in the absence of the lubrication force.  
$(b)$ Bounded motion of two swimmers in the presence of the lubrication force. 
The corresponding distance between the swimmers as a function of time is shown in  
$(c)$. We set $\lambda = 55/12$ and $\chi = \pi/2$. The initial positions of the swimmers as $(3,3,0)$ (puller) and $(9,9,0)$ (pusher). 
Here, lengths are scaled by radius of the swimmer $a$ and time scaled by $\tau = v/a$.}
\label{fig:lub-bound}
\end{figure}

For the choice, $\chi_1 = \chi_2 = \pi/2$, swimmers move in a plane.
In this case, an isolated swimmer moves in a closed circular path with no net displacement, see fig.~\ref{fig:lub-bound}(a). 
However, the presence of a second swimmer in its proximity changes its movements  dramatically. The hydrodynamic forces from the second swimmer convert the two-dimensional circular swimming into three-dimensional helical swimming, 
see fig.~\ref{fig:lub-bound}(b,c). Though in some situations, the pair of swimmers perform a bounded motion, however, in the other situations, they drift away from each other in the long time limit (see fig.~\ref{fig:lam-chi}). 
These behaviors are  therefore sensitive to the strength of the flow fields 
$(\lambda)$ and the associated lubrication force. 
Note that the origin of the bounded motion, in this situation, is the lubrication force acting between the swimmers. 
However, the bounded motion is less stable here, and in some cases, the swimmers diverge from each other in drifting circular paths. Note that, for other values of $\chi$, the bounded motion occurs due to the helical propulsion of the swimmers, and $\chi$ plays a more crucial role than $\lambda$ there. 
Consequently, the former bounded motion is more stable compared to the ones observed due to the lubrication forces.

Notably, a bound state was observed experimentally for a pair of spinning bottom-heavy \textit{Volvox} due to the combined interface effect, gravity, and lubrication forces \cite{drescher,Goldstein_JFM1, Goldstein_JFM2}. 
Here, the bound state is observed for three-dimensional chiral swimmers due to far--field hydrodynamic interaction among them. Note that the bounded motion is restricted to parallel swimming with equal strength of flow field of chiral swimmers, i.e., swimmers with identical $\chi$ and $\lambda$, see figs.~\ref{fig:lam-chi} \& \ref{fig:states}. However, for certain situations, say, $\chi_1 = \chi_2 = \pi/2$, we encounter bound states originating from the combined effect of lubrication force and hydrodynamic attraction.

\section{Influence of the initial configuration of the swimmers}
\label{sec:ini_pos}

\begin{figure}[hbt]
\centering
\includegraphics[scale=1]{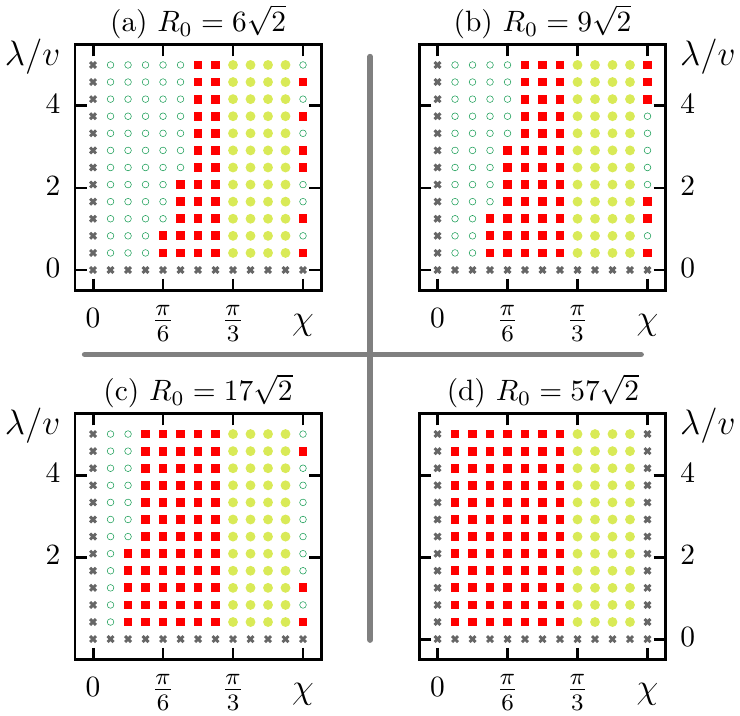}
\caption{Numerically obtained swimming behaviors of pusher-puller type chiral swimmers 
with different initial positions. The corresponding initial distance is $R_0$.  
(a) For ${\bf q}_1 = (9,9,0) a$ and ${\bf q}_2 = (3,3,0) a$ (as in Fig.~\ref{fig:lam-chi}(c)), 
(b) for ${\bf q}_1 = (12,12,0) a$ and ${\bf q}_2 = (3,3,0) a$, 
(b) for ${\bf q}_1 = (20,20,0) a$ and ${\bf q}_2 = (3,3,0) a$, and
(b) for ${\bf q}_1 = (60,60,0) a$ and ${\bf q}_2 = (3,3,0) a$.
Swimmers have the same 
initial velocity ${\bf V}_1 = {\bf V}_2 = v (0, 0, 1)$ 
and rotation rate ${\bf \Omega}_1 = {\bf \Omega}_2 = v\,(\cos \chi, 0, \sin \chi)/a$, 
which depends on the angle $\chi$.
Symbols are same as in Fig.~\ref{fig:lam-chi}.}
\label{fig:ini_pos}
\end{figure}

In this section, we study the impact of the initial distance $R_0$ between the swimmers on their hydrodymic behavior. 
As a test case, we consider the pusher-puller combination, see fig.~\ref{fig:ini_pos}. 
The nature of hydrodynamic interaction between the swimmers changes with varying $R_0$.
As $R_0$ increases swimmers exhibit mainly B, D, and MD states. 
Due to the lubrication forces, the states C or MC do not appear in the state diagrams. 
As $R_0$ increases swimmers tend to exhibit B states than MD.
The general tendency of the swimmers is repulsive or attractive. 
For purely repulsive situation swimmers exhibit MD state. 
If swimmers tend to exhibit attractive behaviour then based on their near field interactions the swimming behavior can be classified as D or B. 
With increasing $R_0$, the flow field of the swimmers prohibit them to approach close to each other. Thus, swimmers exhibit bounded states only.
Note that the lubrication forces become redundant for higher $R_0$ values. 
For $\chi = \pi/2$ swimmers exhibit bounded states between the MD states, depending on strength of $\lambda$, at lower $R_0$ values.  
However, as $R_0$ increases, these D states are converted in to B states.   
In the other combination of swimmers, e.g., pusher-pusher or puller-puller, 
MD states do not alter with respect to $R_0$.
However, for smaller $R_0$, swimmers mostly remain in D state. With increasing $R_0$ (intermediate region), the probability that the swimmers will be bounded to each other increases (see fig.~\ref{fig:ini_pos}).
If $R_0$ is very high $(\sim 10^3)$, the swimmers never approach each other very close so that they cannot interact effectively. Swimmers moving in straight lines or having no stresslet do not interact with each other also (gray cross states in the state diagrams).
Note that for $R_0 \sim 10^3$, the hydrodynamic interaction becomes ineffective.

Note that, as reported in our earlier work \cite{burada}, 
the bound state is stable even with a small perturbation to their initial orientation, say, $(-0.006\pi/24) \leq \psi_1 \leq (0.007\pi/24)$, $(-0.007\pi/24) \leq \psi_2 \leq (0.007\pi/24)$, and $(-1.4\pi/24) \leq \psi_3 \leq (\pi/24)$. Beyond this range the B states are converted into divergence states. Here, $\psi_1, \psi_2$ and $\psi_3$ are initial rotations about $\mathbf{t}_2$, $\mathbf{b}_2$ and $\mathbf{n}_2$ axes respectively. Notably, $(\mathbf{n}_1, \mathbf{b}_1, \mathbf{t}_1)$ and $(\mathbf{n}_2, \mathbf{b}_2, \mathbf{t}_2)$ are material frame of references of the first and second swimmers. While the first swimmer is initially aligned along the $z$-axis, the initial orientation of the second swimmer is perturbed by $(\psi_1, \psi_2, \psi_3)$. Note that D and MD states are not influenced by the initial perturbation in orientation of the swimmers.

\section{Conclusions}
\label{sec:conclusions}

In this paper, we have determined the near-field interaction between the two chiral swimmers using the lubrication theory.
The hydrodynamic force and the torque on a swimmer due to the presence of other swimmer  have been determined analytically, in the lubrication region, and deployed in the numerical simulations to investigate the hydrodynamic interaction between the two swimmers. When the swimmers approach very close to each other, the lubrication force drives the swimmers away from each other in the long time limit. 
Consequently, due to near and far-field hydrodymic interactions two chiral swimmers 
exhibit only monotonic divergence, divergence, and bounded states. 
We find that the coupling of near and far-field hydrodynamic interactions convert the planar circular movement of a swimmer, observed for $\chi = \pi/2$, into three-dimensional helical swimming. 
This leads to an unstable bounded motion of a pair of swimmers.  
However, the stable bounded motion of the swimmers, observed for $\chi < \pi/2$, is solely due to the far-field hydrodynamic interaction between the swimmers. 
This study is useful to understand the collective behavior of ciliated microorganisms and artificial swimmers \cite{paxton,ismagilov,golestanianart,dreyfus,hogg}.

\section*{Acknowledgements}

This work was supported by the Indian Institute of Technology Kharagpur, India.

\appendix

\section{Lubrication force}
\label{lub_cal}

\begin{figure}[h]
\centering
\includegraphics[scale=0.65]{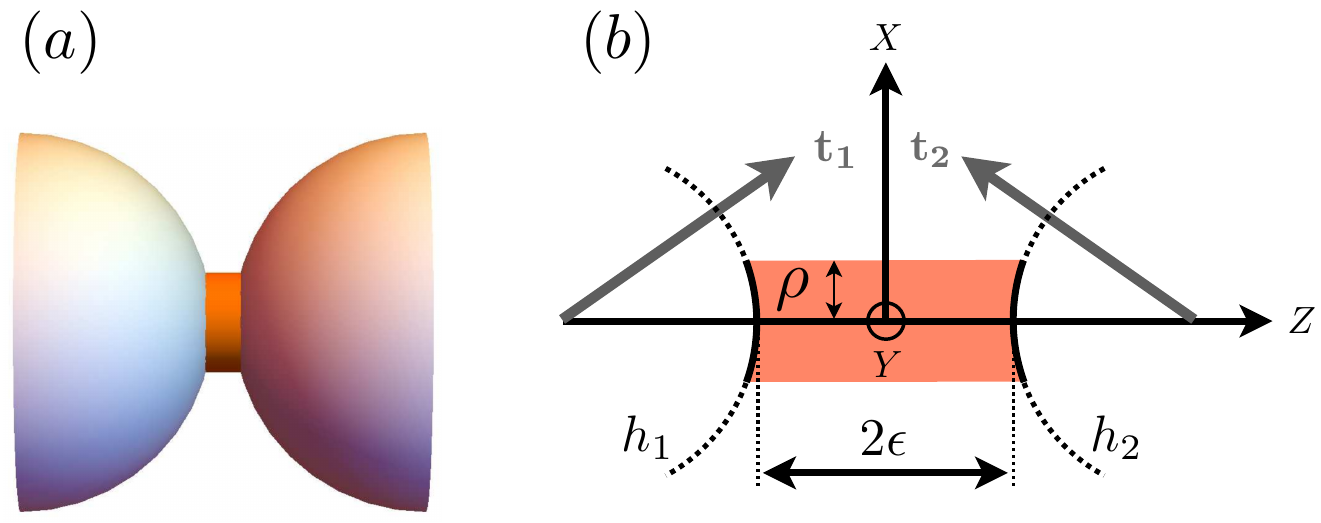}
\caption{
(a) schematic diagram of the lubrication region. 
$(b)$ schematic of the cylindrical region of length $2 \epsilon$ between the spherical swimmers. Here, $\rho$ is the radius of the cylinder, 
$X$, $Y$, and $Z$ form the cartesian frame whose origin is at the middle of the parabolic surfaces $h_1$ and $h_2$ ($h_2 = -h_1$) of swimmer one and two, respectively. 
The corresponding radial vector is defined as $\bm{\rho} = X\mathbf{e}_X + Y \mathbf{e}_Y = \rho \, \mathbf{e}_\rho$, where $\mathbf{e}_\rho$ is the unit radial vector, and 
$\mathbf{e}_\phi$ is the unit vector along the azimuthal direction (on XY-plane) in the cylindrical region.
$\mathbf{t_1}$ and $\mathbf{t_2}$ are the orientations of the swimmers.}
\label{lub_cala}
\end{figure}

We briefly explain here the lubrication calculations \cite{wang}. 
When the spherical swimmers approach each other, i.e., $R < 2(a + \epsilon)$, the narrow gap between them forms a cylindrical region (see fig.~\ref{lub_cala}). 
Here, $R$ is the distance between the swimmer, $a$ is radius of the swimmer, and $\epsilon$ is half of the distance between the swimmers.
The flow fields generated by the swimmers obey the Stokes equation, Eq.~\ref{eq:stokes}, 
in this region.
The surfaces of the two spherical swimmers in the narrow gap region can be considered as parabolic surfaces having the form, 
\begin{align}
h_1 &= \epsilon + \frac{\rho^\prime\,{^2}}{2} + ...\ , \\
h_2 &= - h_1 \,,
\end{align}
where $\rho^\prime$ is the dimensionless radius in cylindrical coordinates. 
We set the origin is at the mid point between two spherical squirmers. 
The stretched coordinates ($X, Y, Z$) \cite{wang} used here are defined as (see fig.~\ref{lub_cala}(b)),
\begin{equation}
\begin{split}
\sqrt{\epsilon} X &= x \ , \sqrt{\epsilon} Y = y \ , \epsilon Z = z \ , \\ 
 \rho^\prime &= \sqrt{x^2 + y^2}   \, , 
 \sqrt{\epsilon} \rho = \rho^\prime.
\end{split}
\end{equation} 
Accordingly, the scaled surfaces are defined as 
$H_1 = h_1/\epsilon$ and $H_2 = h_2/\epsilon = -H_1$. 
The radial vector in the stretched coordinates is defined as $\bm{\rho} = X\mathbf{e}_X + Y \mathbf{e}_Y = \rho \, \mathbf{e}_\rho$, where $\mathbf{e}_\rho$ is the unit radial vector.
The Stokes equation, Eq.~\ref{eq:stokes}, in the stretched coordinates can be expressed in dimensionless form as, 
\begin{align}
\left[ \epsilon \left( \frac{\partial^2}{\partial X^2} + \frac{\partial^2}{\partial Y^2} \right) + \frac{\partial^2}{\partial Z^2}\right] \mathbf{u} &= \epsilon \left( \epsilon^{1/2} \frac{\partial p}{\partial X},  \epsilon^{1/2} \frac{\partial p}{\partial Y}, \frac{\partial p}{\partial Z} \right) \, , \\
\epsilon^{1/2} \left( \frac{\partial u}{\partial X} + \frac{\partial v}{\partial Y} \right) + \frac{\partial w}{\partial Z} &= 0 \,
\end{align}
where $u, v$, and $w$ are the components of the velocity field, and 
$p$ is the pressure field.

The surface slip (Eq.~\ref{eq:slip}) of swimmer one, for $l = 1$ mode, is given by,
\begin{align}
\mathbf{u^{s1}} = -\beta_{10}^1 [(\mathbf{t_1} \cdot \mathbf{e_r})\, \mathbf{e_r} - \mathbf{t_1}] - (\mathbf{t_1} \times \mathbf{e}_r)\gamma_{10}^1 \,,
\end{align}
where $\mathbf{t}_1$ is the swimming direction and $\mathbf{e}_r$ is the unit radial vector measured from center of the swimmer one. 
Similarly, for swimmer two, 
\begin{align}
\mathbf{u^{s2}} = -\beta_{10}^2[(\mathbf{t}_2 \cdot \mathbf{e}_r')\,\mathbf{e}_r' - \mathbf{t_2}] - (\mathbf{t}_2 \times \mathbf{e}_r')\gamma_{10}^2 \ ,
\end{align}
where $\mathbf{t}_2$ is the swimming direction and $\mathbf{e}_r'$ is the unit radial vector measured from center of the swimmer two. 
Following the procedure given by Ishikawa \textit{et. al.} \cite{simmonds}, 
we expand the velocity and pressure fields on the surface of the swimmer 
in terms of $\epsilon^{1/2}$ as, 
\begin{align}
\label{eq:vp1}
\mathbf{u^s} &= \mathbf{u_0^s} + \epsilon^{1/2} \, \mathbf{u_1^s} + ... \ , \\
\label{eq:vp2}
p &= p_\infty + \epsilon^{-3/2} \, (p_0 + \epsilon^{1/2}p_1 + ...) \ .
\end{align}
Similarly, the surface slip of swimmer one and two can be expanded in terms of $\epsilon^{1/2}$ as 
$\mathbf{u^{s1}} = \mathbf{u_0^{s1}} + \epsilon^{1/2} \, \mathbf{u_1^{s1}} + ...$ and 
$\mathbf{u^{s2}} = \mathbf{u_0^{s2}} + \epsilon^{1/2} \, \mathbf{u_1^{s2}}+ ...$, respectively, 
where
\begin{align}
\mathbf{u_0}^{s1} &= \beta_{10}^1 \, [(\mathbf{t_1} \cdot \mathbf{e_z}) \, \mathbf{e_z} - \mathbf{t_1}] + (\mathbf{t_1} \times \mathbf{e_z})\,\gamma_{10}^1 \ , \\
\mathbf{u_1}^{s1} &= \beta_{10}^1 \, [(\mathbf{t_1} \cdot \bm{\rho}) \, \mathbf{e_z} + (\mathbf{t_1} \cdot \mathbf{e_z}) \, \bm{\rho}] - (\mathbf{t_1}\times \bm{\rho}) \, \gamma_{10}^1 \ , \\
\mathbf{u_0}^{s2} &= -\beta_{10}^2 \, [(\mathbf{t_2} \cdot \mathbf{e_z}) \, \mathbf{e_z} - \mathbf{t_2}] - (\mathbf{t_2} \times \mathbf{e_z}) \, \gamma_{10}^2 \ , \\
\mathbf{u_1}^{s2} &= \beta_{10}^2 \, [(\mathbf{t_2} \cdot \bm{\rho}) \, \mathbf{e_z} + (\mathbf{t_2} \cdot \mathbf{e_z}) \, \bm{\rho}] - (\mathbf{t_2}\times \bm{\rho}) \, \gamma_{10}^2 \ .
\end{align}
Here, 
$\mathbf{t_1} = t_{11} \,\mathbf{e_X} + t_{12} \,\mathbf{e_Y} + t_{13} \,\mathbf{e_Z}$ and 
$\mathbf{t_2} = t_{21} \mathbf{e_X} + t_{22} \mathbf{e_Y} + t_{23} \mathbf{e_Z}$. Note that $\mathbf{e}_X, \mathbf{e}_Y$, and  $\mathbf{e}_Z$ are the unit vectors along the stretched coordinates $X, Y$ and $Z$, respectively. 

Following the procedure by Wang \textit{et. al} \cite{wang}, we get the solutions for the velocity and pressure fields in the lubrication region. 
We found that, in the lubrication region, only the first order term survives  
in the solution of the pressure field, and the contribution from the other terms is negligible in the limit $\epsilon \to 0$. 
To the first order, the lubrication equations for the given system are,
\begin{subequations}
\label{eq:eq1}
\begin{align}
\frac{\partial p_1}{\partial X} &= \frac{\partial^2 u_1}{\partial Z^2}   \, \\
\frac{\partial p_1}{\partial Y} &= \frac{\partial^2 v_1}{\partial Z^2}   \ , \\
\frac{\partial p_1}{\partial Z} &= 0 \,,
\end{align} 
\end{subequations}
where $u_1, v_1$, and $p_1$ are the components of the velocity and pressure fields, respectively, corresponding to the first terms (see Eqs.~\ref{eq:vp1} \& \ref{eq:vp2}).

As the velocity field is equal to the active slip at the surface of the swimmer, 
the corresponding components (first order) of the surface slip of swimmer one read, 
\begin{subequations}
\begin{align}
u_{11} &= \mathbf{u_1}^{s1} \cdot \mathbf{e}_X = \beta_{10}^1t_{13}X + \gamma_{10}^1t_{13}Y \, , \\ 
v_{11} &= \mathbf{u_1}^{s1} \cdot \mathbf{e}_Y = \beta_{10}^1t_{13}Y - \gamma_{10}^1t_{13}X \, , \\
w_{11} &= \mathbf{u_1}^{s1} \cdot \mathbf{e}_Z = \beta_{10}^1(\mathbf{t}_1\cdot\mathbf{e}_\rho)\rho + \gamma_{10}^1 \rho (\mathbf{t}_1\cdot\mathbf{e}_\phi) \ , 
\end{align}
\end{subequations}
where $\mathbf{e}_\phi$ is the unit vector along the azimuthal direction (on XY-plane) in the cylindrical region. 
Similarly, for the components of the surface slip of swimmer two read, 
\begin{subequations}
\begin{align}
u_{21} & = \mathbf{u_1}^{s2} \cdot \mathbf{e}_X = -\beta_{10}^2t_{23}X + \gamma_{10}^2t_{23}Y \, , \\ 
v_{21} & = \mathbf{u_1}^{s2} \cdot \mathbf{e}_Y = -\beta_{10}^2t_{23}Y - \gamma_{10}^2t_{23}X \, , \\
w_{21} & = \mathbf{u_1}^{s2} \cdot \mathbf{e}_Z = -\beta_{10}^2 (\mathbf{t}_2\cdot\mathbf{e}_\rho)\rho + \gamma_{10}^2 \rho (\mathbf{t}_2\cdot\mathbf{e}_\phi)\,.
\end{align}
\end{subequations}
Note that, in laboratory frame of reference, 
the velocity field is zero, i.e., $\mathbf{u} = 0$ far away from the swimmers.  
Integrating Eq.~\ref{eq:eq1} twice we get,
\begin{subequations}
\begin{align}
u_1 &= \frac{Z^2-H_1^2}{2} \, \frac{\partial p_1}{\partial X} + \frac{Z}{2H_1}(u_{11}-u_{21})+\frac{1}{2}(u_{11}+u_{21}) \ \label{u1}\,, \\
v_1 &= \frac{Z^2-H_1^2}{2} \, \frac{\partial p_1}{\partial Y} + \frac{Z}{2H_1}(v_{11}-v_{21})+\frac{1}{2}(v_{11}+ v_{21}) \ \label{v1} \,.
\end{align}
\end{subequations}
Now, differentiating Eq.~\ref{u1} with respect to $X$ and Eq.~\ref{v1} with respect to 
$Y$ we get,
\begin{subequations}
\begin{align}
\label{eq:D_u_x}
\frac{\partial u_1}{\partial X} & = 
\frac{Z^2-H_1^2}{2}\,\frac{\partial^2 p_1}{\partial X^2} - H_1X\frac{\partial p_1}{\partial X}
+ \frac{1}{2}(\beta_{10}^1t_{13} - \beta_{10}^2t_{23}) \nonumber \\ 
& +\frac{Z}{(2 + X^2 + Y^2)^2}
 \Big[-(\beta_{10}^1 t_{13} + \beta_{10}^2 t_{23})(-2 + X^2) \nonumber \\
& + 2(-\gamma_{10}^1 t_{13} + \gamma_{10}^2 t_{23})XY  
 + Y^2 (\beta_{10}^1 t_{13} + \beta_{10}^2 t_{23})  \Big] \,,
\end{align}
\begin{align}
\label{eq:D_u_y}
\frac{\partial v_1}{\partial Y} & = 
\frac{Z^2-H_1^2}{2}\,\frac{\partial^2 p_1}{\partial Y^2} - H_1Y\frac{\partial p_1}{\partial Y} + \frac{1}{2}(\beta_{10}^1t_{13} - \beta_{10}^2t_{23}) \nonumber \\  
&+ \frac{Z}{(2 + X^2 + Y^2)^2}\Big[ (\beta_{10}^1 t_{13} + \beta_{10}^2 t_{23})(2 + X^2) \nonumber \\
&  + 2(\gamma_{10}^1 t_{13} - \gamma_{10}^2 t_{23})XY  
 - Y^2 (\beta_{10}^1 t_{13} + \beta_{10}^2 t_{23}) \Big]\,.
\end{align}
\end{subequations}
Adding Eq.~\ref{eq:D_u_x} and Eq.~\ref{eq:D_u_y} we get,
\begin{align}
\frac{\partial u_1}{\partial X} + \frac{\partial v_1}{\partial Y} 
 = & \frac{Z^2- H_1^2}{2} \,\nabla^2p_1  - H_1(\bm{\rho}\cdot \nabla)p_1 \nonumber \\
& + B + \frac{4 Z}{(2 + X^2 + Y^2)^2}D_1\ \label{eq_cont} \,,
\end{align}
where
$B = \beta_{10}^1 t_{13} - \beta_{10}^2 t_{23}$ and 
$D_1 = \beta_{10}^1 t_{13} + \beta_{10}^2 t_{23}$.
Integrating Eq.~\ref{eq_cont} with respect to $Z$ between the two surfaces, and 
using the incompressibility condition we get,
\begin{align}
\int_{H_1}^{H_2} \Big( \frac{\partial u_1}{\partial X} + \frac{\partial v_1}{\partial Y} \Big)dZ 
& = - \int_{H_1}^{H_2} \frac{\partial w_1}{\partial Z} dZ \, \\
\frac{2H_1^3}{3}\nabla^2p_1 + 2H_1^2(\bm{\rho}\cdot\nabla) \, p_1 - 2BH_1 
& = \rho \, (\mathbf{E}_{12}\cdot\mathbf{e}_\rho + \mathbf{E}_{12}'\cdot\mathbf{e}_\phi)\,,
\end{align}
where $\mathbf{E}_{12}= \beta_{10}^1\mathbf{t}_1 + \beta_{10}^2\mathbf{t}_2$ and 
$\mathbf{E}_{12}'= \gamma_{10}^1\mathbf{t}_1 - \gamma_{10}^2\mathbf{t}_2$.
Using the operators, $\nabla^2$ and $\nabla$, in cylindrical coordinate system 
the above equation can be simplified as
\begin{align}
\label{eq_pressure}
& \frac{2 H_1^3}{3}\Big[ \frac{1}{\rho}\frac{\partial}{\partial \rho}\Big(\rho \frac{\partial}{\partial \rho}\Big) + \frac{1}{\rho^2}\frac{\partial^2}{\partial \phi^2}\Big]p_1 + 2 H_1^2\rho \frac{\partial p_1}{\partial \rho} - 2BH_1 \nonumber \\ 
& = \rho\, (\mathbf{E}_{12}\cdot\mathbf{e}_\rho + \mathbf{E}_{12}'\cdot\mathbf{e}_\phi)\,. 
\end{align}
Note that the pressure term does not contains $Z$ component.
Let,
\begin{align}
p_1 &= p_a + p_s (\mathbf{E}_{12}\cdot\mathbf{e}_\rho)+ p_m (\mathbf{E}_{12}'\cdot\mathbf{e}_\phi)\,, 
\label{eq_p1}
\end{align}
where $p_a$, $p_s$ and $p_m$ are the solutions of Eq.~(\ref{eq_pressure}). 
Inserting Eq.~\ref{eq_p1} in Eq.~\ref{eq_pressure} we get the equation for the 
particular solution as,
\begin{align}
\frac{ H_1^2}{3 \rho} \, \frac{\partial}{\partial \rho}\Big(\rho \frac{\partial p_a}{\partial \rho}\Big) +  H_1\rho \frac{\partial p_a}{\partial \rho} - B &= 0 \ .
\end{align}
This gives us, 
\begin{align}
 \label{p_a}
p_a(\rho) &= - B \Big[ \frac{3}{4 H_1} + \frac{3}{8 H_1^2} \Big] \,.
\end{align}
Note that $p_a$ has no $\phi$ dependency. The 2nd term in Eq.~\ref{eq_p1} ($p_s$) gives us,
\begin{align}
\label{eq:ps}
& \frac{2 H_1^3}{3}\Big[ \frac{1}{\rho}\frac{\partial}{\partial \rho}\Big(\rho \frac{\partial}{\partial \rho}\Big) + \frac{1}{\rho^2}\frac{\partial^2}{\partial \phi^2}\Big]p_s (\mathbf{E}_{12}\cdot\mathbf{e}_\rho) \nonumber \\
 & + 2 H_1^2\rho \frac{\partial }{\partial \rho} p_s (\mathbf{E}_{12}\cdot\mathbf{e}_\rho) 
 = \rho (\mathbf{E}_{12}\cdot\mathbf{e}_\rho) \ .
\end{align}
Using the relations, 
$\partial \mathbf{e}_\rho/\partial \rho = 0$, 
$\mathbf{e}_\phi = \partial \mathbf{e}_\rho/\partial \phi$, and
$\mathbf{e}_\rho = -\partial \mathbf{e}_\phi/\partial \phi$, 
Eq.~\ref{eq:ps} can be simplified as,
\begin{align}
\label{eq:ps_1}
\frac{2 H_1^3}{3 \, \rho}\frac{\partial}{\partial \rho}\Big(\rho \frac{\partial p_s}{\partial \rho}\Big) - \frac{2 H_1^3}{3 \rho^2}p_s+2 H_1^2\rho \frac{\partial p_s}{\partial \rho} - \rho &= 0 \,.
\end{align}
The solution of Eq.~\ref{eq:ps_1} is given by,
\begin{align}
\label{eq:ps_2}
p_s(\rho) = - \frac{6 \rho}{5 (2 + \rho^2)^2} \ .
\end{align}
One can follow a similar procedure to obtain the solution for $p_m(\rho)$ as
\begin{align}
\label{eq:pm_2}
p_m(\rho) =  - \frac{6 \rho}{5 (2 + \rho^2)^2} \,.
\end{align}
Incidentally, the solutions of $p_s$ and $p_m$ are the same.  
Therefore, from Eqs.~\ref{p_a}, ~\ref{eq:ps_2}, and ~\ref{eq:pm_2} we get the solution 
for $p_1$ as,
\begin{align}
\label{eq:p1}
p_1 = & - B \Big[ \frac{3}{4 H_1} + \frac{3}{8 H_1^2} \Big] - \frac{6\rho}{5 (2 + \rho^2)^2} \big[ \nonumber \\
&(\beta_{10}^1 \mathbf{t}_1 + \beta_{10}^2 \mathbf{t}_2) \cdot \mathbf{e}_\rho +
(\gamma_{10}^1 \mathbf{t}_1 - \gamma_{10}^2 \mathbf{t}_2) \cdot \mathbf{e}_\phi
\big]
\end{align} 
The corresponding velocity field (Eqs. ~\ref{u1} \& ~\ref{v1}) of 
swimmer one can be determined in the lubrication region as,   
\begin{subequations}
\begin{align}
u_{\rho,1} &= \frac{3 (Z^2 - H_1^2)}{10(2 + \rho^2)^3} \Big[ 5B\rho^3 + 20 B \rho - 4(\mathbf{e}_\rho \cdot \mathbf{E}_{12} + \mathbf{e}_\phi \cdot \mathbf{E}'_{12}) \nonumber \\ &+ 6 \rho^2 (\mathbf{e}_\rho \cdot \mathbf{E}_{12} + \mathbf{e}_\phi \cdot \mathbf{E}'_{12}) \Big] + \frac{Z \rho}{2H_1}D_1 
+ \frac{\rho}{2}B, 
\end{align}
\begin{align}
u_{\phi,1} &=  \frac{6(Z^2 - H_1^2)}{10(2 + \rho^2)^2} \Big[
- \mathbf{e}_\phi \cdot \mathbf{E}_{12} + \mathbf{e}_\rho \cdot \mathbf{E}'_{12} \Big] 
+ \frac{Z \rho}{2 H_1}[\gamma_{10}^2 t_{23} \nonumber \\
& - \gamma_{10}^1 t_{13}] - \frac{\rho}{2}[\gamma_{10}^2 t_{23} + \gamma_{10}^1 t_{13}]
\,,
\end{align}
\begin{align}
u_{Z,1} & = \frac{Z}{20 (2 + \rho^2)^4} 
\Big[ 
-5 B ((2 + \rho^2)^2 - 4 Z^2) \cdot (-8 + \nonumber \\
& 4 \rho^2 + \rho^4) +
 12 (2 + \rho^2)^2(6 + \rho^2)\Big(X (A_1 + C_2) \nonumber \\ 
 & + Y(-A_2 + C_1)\Big)
 - 40 D_1 Z (2 + \rho^2)^2 + \nonumber \\
& 32(-4 + \rho^2)\Big(X(A_1 + C_2)+ (-A_2 + C_1)Y\Big)Z^2 \Big] \nonumber \\
& + \frac{1}{2} \Big[ D_1 + X(-\beta_{10}^2 t_{21} + \gamma_{10}^1 t_{12} + \gamma_{10}^2 t_{22}) \nonumber \\
& - Y(\beta_{10}^2 t_{22} + \gamma_{10}^1 t_{11} + \gamma_{10}^2 t_{21}) + \beta_{10}^1 (t_{11} X + t_{12}Y) \Big] \,,
\end{align}
\end{subequations}
where
$B = (\beta_{10}^1 t_{13} - \beta_{10}^2 t_{23}), 
D_1 = (\beta_{10}^1 t_{13} + \beta_{10}^2 t_{23}), 
A_1 = (\beta_{10}^1 t_{11} + \beta_{10}^2 t_{21}), 
C_1 = (\beta_{10}^1 t_{12} + \beta_{10}^2 t_{22}), 
A_2 = (\gamma_{10}^1 t_{11} - \gamma_{10}^2 t_{21}),
\,\text{and}\, 
C_2 = (\gamma_{10}^1 t_{12} - \gamma_{10}^2 t_{22})$.

Finally, the force component along the $Z$ direction can be calculated using the relation, $dF_{_Z} = \mathbf{e}_Z \cdot (\mathbf{\sigma_1} \cdot \mathbf{n}_1)\, dA$, where 
$\mathbf{\sigma_1}$ is the corresponding stress tensor, 
$dA$ is the area element on the swimmer surface and 
$\mathbf{n}_1 = -\cos\theta \mathbf{e}_Z + \sin\theta \mathbf{e}_\rho$ (normal vector, see fig.~\ref{fig:sketch1}).
Subsequently, we can calculate the force component as,
\begin{align}
F_{_Z} & = - \frac{3 \pi B a^2}{2} 
\left[- \ln(2) + \ln (2 + \rho_0^2)\right]\,.
\label{eq:lub_force_A}
\end{align}
Here, $\rho_0$ is the distance up to which the lubrication force is considerable. Generally, $\rho_0 = a \epsilon^{-1}$. Therefore,
\begin{align}
F_{_Z} & \approx \frac{3 \pi B a^2}{2} \ln \epsilon \, .
\end{align}

The corresponding toque along the $Y$-direction can be calculated using the relation, 
$d T_{_Y} = - (\mathbf{n}_1 \cdot \mathbf{e}_{_X}) dF_{_Z} 
+ (\mathbf{n}_1 \cdot \mathbf{e}_{_Z}) dF_{_X}$. 
The torque is given by, 
\begin{align}
T_{_Y} &= \epsilon^{1/2} \, \frac{3\pi}{10} \Big[ (\beta_{10}^1 t_{11} + \beta_{10}^2 t_{21}) + (\gamma_{10}^1 t_{11} - \gamma_{10}^2 t_{21})\Big] \nonumber \\
& \Big[ \frac{8}{2 + \rho^2} + 3 \ln(2 + \rho^2)+ 4 + \ln\, 8\Big] \, .
\end{align}
Similar expression for torque can be obtained about the $X$-direction as well. 
However note that torques are of the order $\epsilon^{1/2}$, and 
the contribution of the torques to the rotational motion of the swimmers is negligible. 
Thus, we do not include them in the numerical simulations. 

\section{$\chi - \chi$ and $\lambda - \lambda$ state diagrams}
\label{state}

\begin{figure*}[htb!]
\includegraphics[scale=1]{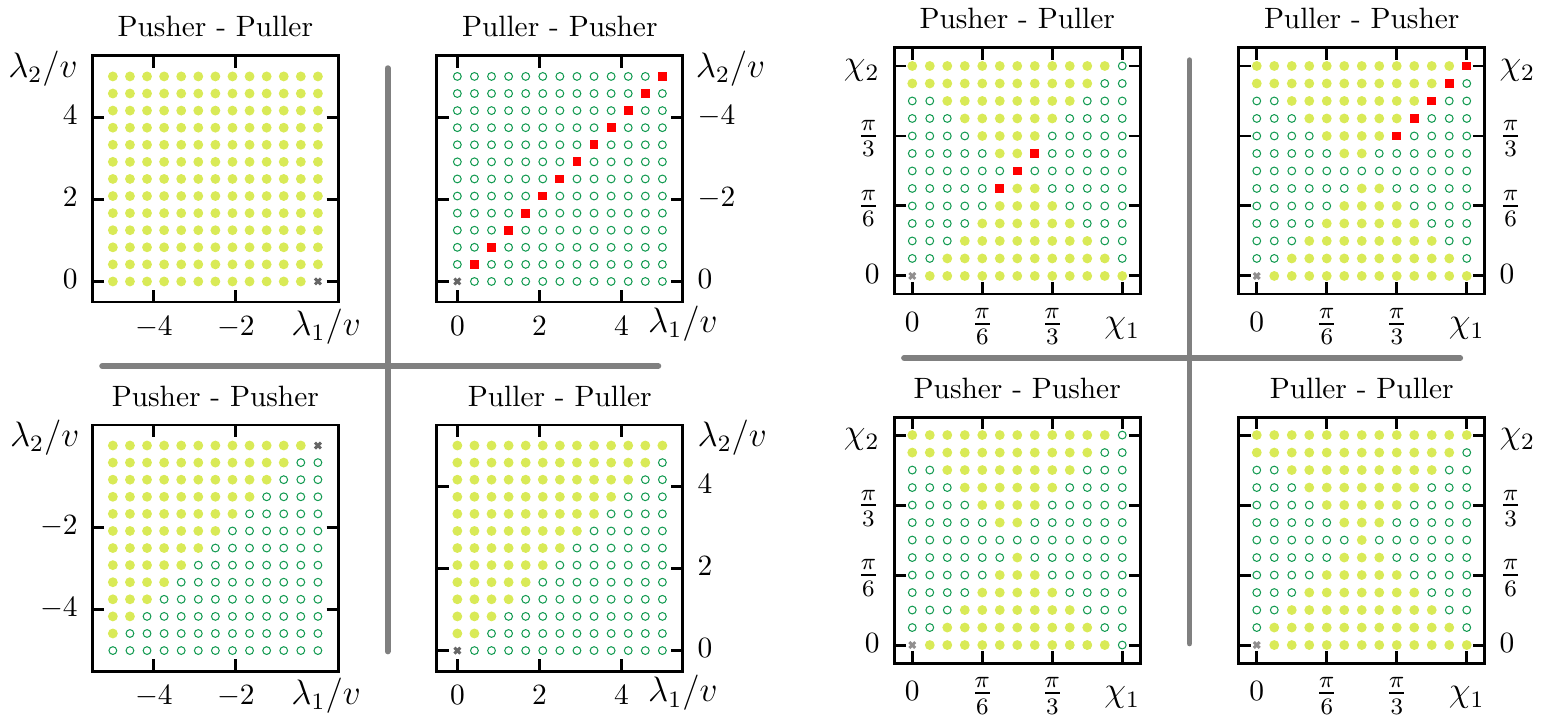}
\caption{State diagrams generated for varying hydrodynamic field strengths
$(\lambda_1, \lambda_2)$, at fixed $\chi_1 = \chi_2 = \chi$, and $(\chi_1, \chi_2)$, at fixed $\lambda_1 = \lambda_2 = \lambda$. 
Initial conditions and color codes are given in fig. \ref{fig:lam-chi}.}
\label{fig:states}
\end{figure*}

Aforementioned, a pair of chiral swimmers exhibit mainly B, D, and MD states in the presence of the lubrication forces. Fig.~\ref{fig:lam-chi} depicted these states 
for the choice $\lambda_1 = \lambda_2 = \lambda$ and $\chi_1 = \chi_2 = \chi$. 
However, one can also vary the parameters $\lambda_1,\lambda_2,\chi_1$, and $\chi_2$ 
to study the hydrodynamic behavior of two chiral swimmers (see fig.~\ref{fig:states}).
With varying, $\lambda_1,\lambda_2,\chi_1$, and $\chi_2$, swimmers mainly exhibit the D and MD states. Only in the asymmetric combination of pusher and puller type swimmers exhibit bounded states for $|\lambda_1| = |\lambda_2|$ or $\chi_1 = \chi_2$. 
This means, swimmers with same $\mathbf{V}$ and $\mathbf{\Omega}$, and equal strength of flow field (however, the sign of $\lambda$ should be different) exhibit the interesting bounded states. 
Note that when the swimmers are very close to each other, due to the lubrication force,  swimmers repel each other strongly and they do not exhibit either convergence (C) or monotonic convergence (MC) states. A similar behavior can be observed in the case of 
axisymmetric squirmers.

\end{document}